\documentclass
[superscriptaddress,secnumarabic,amssymb,amsmath,nobibnotes,aps,prd,showkeys,showpacs,nofootinbib,onecolumn,12pt]{revtex4}%
\usepackage{graphicx}
\usepackage{epsf}
\usepackage{bm}
\usepackage{amsmath}
\usepackage{amsfonts}
\usepackage{amssymb}%
\providecommand{\U}[1]{\protect\rule{.1in}{.1in}}

\newcommand{\be}{\begin{equation}}
\newcommand{\ee}{\end{equation}}

\newcommand{\mincir}{\raise
-3.truept\hbox{\rlap{\hbox{$\sim$}}\raise4.truept\hbox{$<$}\ }}
\newcommand{\magcir}{\raise
-3.truept\hbox{\rlap{\hbox{$\sim$}}\raise4.truept\hbox{$>$}\ }}

\begin{document}
\title{New Anisotropic Exact Solution in Multifield Cosmology}
\author{Andronikos Paliathanasis}
\email{anpaliat@phys.uoa.gr}
\affiliation{Institute of Systems Science, Durban University of Technology, Durban 4000,
South Africa}
\affiliation{Instituto de Ciencias F\'{\i}sicas y Matem\'{a}ticas, Universidad Austral de
Chile, Valdivia, Chile}

\begin{abstract}
In the case of two-scalar field cosmology and, specifically for the Chiral
model, we determine an exact solution for the field equations with a
anisotropic background space. The exact solution can describe anisotropic
inflation with a Kantowski-Sachs geometry and it can be seen as the
anisotropic analogue of the hyperbolic inflation. Finally, we investigate the
evolution for the homogeneous perturbations around the new asymptotic solution.

\end{abstract}
\keywords{Multifield Cosmology; Anisotropic spacetimes; Kantowski-Sachs; Anisotropic inflation}
\pacs{98.80.-k, 95.35.+d, 95.36.+x}
\date{\today}
\maketitle

\section{Introduction}

\label{sec1}

The early acceleration epoch of the universe is the inflationary era
\cite{guth}, to which the isotropy and homogeneity of the observed universe
are due \cite{inf1}. The origin of the inflation is unknown. However, the
introduction of a minimally coupled scalar field, the inflaton, into the
cosmological dynamics of Einstein's General Relativity provides an
acceleration when the scalar field potential dominates. Hence, the scalar
field drives the spacetime towards a locally isotropic and homogeneous space
form that leaves only very small residual anisotropies, which are left from
the pre-inflationary era \cite{inf2,inf3}. Therefore, anisotropies may have
been important for the evolution of the universe. Thus, the investigation of
exact solutions in anisotropic inflationary models is a subject of special interest.

Exact and analytic solutions are important for the study of the evolution and
of the viability of a given cosmological model. In one scalar field cosmology,
exact and analytic solutions in a homogeneous and isotropic background space
can be found in \cite{sf1,sf2,sf3,sf4,sf5,sf6,sf7,sf8,sf9,sf10}. On the other
hand, there are few known anisotropic exact solutions with one scalar field
\cite{sf11,sf12,sf13,sf14,sf15,sf16}.

Multiscalar field models have been proposed as alternative models for the
description of the whole cosmological history \cite{mult1,quin}. In the
multiscalar field model the additional degrees of freedom provide new
dynamical behaviours in the cosmological dynamics
\cite{aa1,aa2,aa3,aa4,aa5,aa6,aa7}. Some anisotropic exact solutions in
multifield cosmology can be found in \cite{sf16,sf17,sf18}.

A multiscalar field model which has drawn the attention of cosmologists in
recent years is the Chiral model. The Lagrangian function of the Chiral model
is inspired by the $\sigma$-model \cite{sigm0} and is composed of two scalar
fields and the kinetic energy is defined on a two-dimensional hyperbolic space
\cite{atr6}. The Chiral model with an exponential potential provides a new
inflationary solution known as hyperbolic inflation \cite{vr91,vr92}.
Hyperinflation solves various problems of inflationary physics. In hyperbolic
inflation, the dynamics are driven by all the matter components of the field
equations, that is, by the scalar field potential and the kinetic parts of the
two scalar fields. Moreover, the initial conditions in the start and in the
end of the inflation can be different in the Chiral model, which means that
the curvature perturbations depends upon the number of e-fold \cite{in1}.
Furthermore, detectable non-Gaussianities in the power spectrum are supported
by the multifield inflation \cite{in2}.

In this study we investigate the existence of a new exact solution in Chiral
cosmology with an anisotropic background space. As far as isotropic and
homogeneous models are concerned, Chiral theory has been widely studied
previously with many interesting results, see for instance
\cite{ch1,ch2,ch3,ch4}, while recently extensions of Chiral cosmology were
considered by assuming one of the two scalar fields to be a phantom field
\cite{ch5}. In our consideration for the background space we consider locally
rotational spacetimes (LRS) with two scale factors which belong to the family
of Bianchi I, Bianchi III and Kantowski-Sachs spacetimes. These anisotropic
spacetimes have the property that they fall into the spatially flat, closed
and open Friedmann--Lema\^{\i}tre--Robertson--Walker (FLRW) when they reach
isotropy. The plan of the paper is as follows.

In Section \ref{sec2} we present the cosmological model of our consideration
and we derive the gravitational field equations. In Section \ref{sec3} we
present the new solution of our analysis which is that of anisotropic
hyperbolic inflation. The analysis of homogeneous perturbations is presented
in Section \ref{sec4} where we discuss about the stability properties of the
new exact solution. Finally, in Section \ref{sec5} we draw our conclusions.

\section{Chiral cosmology}

\label{sec2}

We consider the gravitational Action Integral%
\begin{equation}
S=\int\sqrt{-g}dx^{4}\left(  R+L_{C}\left(  \phi,\nabla_{\mu}\phi,\psi
,\nabla_{\mu}\psi\right)  \right)  \label{ch.01}%
\end{equation}
in which $R\left(  x^{\kappa}\right)  $ is the Ricci scalar of the metric
tensor $g_{\mu\nu}\left(  x^{\kappa}\right)  ,$ and $L_{C}\left(  \phi
,\nabla_{\mu}\phi,\psi,\nabla_{\mu}\psi\right)  $ is the Lagrangian function
for the Chiral model which describes the dynamics for the two scalar fields
$\phi\left(  x^{\kappa}\right)  $ and $\psi\left(  x^{\kappa}\right)  $, that
is%
\begin{equation}
L_{C}\left(  \phi,\nabla_{\mu}\phi,\psi,\nabla_{\mu}\psi\right)  =-\frac{1}%
{2}g^{\mu\nu}\left(  x^{\kappa}\right)  \left(  \nabla_{\mu}\phi\left(
x^{\kappa}\right)  \nabla_{\nu}\phi\left(  x^{\kappa}\right)  +e^{-2\kappa
\phi\left(  x^{\kappa}\right)  }\nabla_{\mu}\psi\left(  x^{\kappa}\right)
\nabla_{\nu}\psi\left(  x^{\kappa}\right)  \right)  +V(\phi\left(  x^{\kappa
}\right)  ). \label{ch.02}%
\end{equation}

From the kinetic term of (\ref{ch.02}) we observe that the scalar field lies
on two geometries. The physical space with metric tensor $g_{\mu\nu}\left(
x^{\kappa}\right)  $ and the two-dimensional space of constant curvature with
metric $h_{AB}=diag\left(  1,e^{-2\kappa\phi}\right)  $ and curvature
$R_{h}\simeq-\kappa^{2}$, $A,B=1,2$. The parameter $\kappa$ is assumed to be a
nonzero constant, otherwise the line element $h_{AB}$ reduces to the
two-dimensional flat space and the Lagrangian (\ref{ch.02}) is reduced to that
of multiquintessence theory. In general the potential function (\ref{ch.02})
has been assumed to be also a function of the second field $\psi\left(
x^{k}\right)  $. However, hyperbolic inflation in the case of FLRW space
follows for the exponential potential \cite{vr91} $V\left(  \phi\left(
x^{\kappa}\right)  \right)  =V_{0}\exp\left(  -\lambda\phi\left(  x^{\kappa
}\right)  \right)  $ which we shall consider in this analysis.

\subsection{Anisotropic spacetime}

In this study for the physical space we consider the LRS anisotropic line
element in the Milne variables%
\begin{equation}
ds^{2}=-dt^{2}+e^{2\alpha\left(  t\right)  }\left(  e^{2\beta\left(  t\right)
}dx^{2}+e^{-\beta\left(  t\right)  }\left(  dy^{2}+f^{2}\left(  y\right)
dz^{2}\right)  \right)  \label{ch.03}%
\end{equation}
in which the function $f\left(  y\right)  $ has one of the following forms,
$f_{A}\left(  y\right)  =1$, and the line element describes a Bianchi I
spacetime, $f_{B}\left(  y\right)  =\sinh\left(  \sqrt{\left\vert K\right\vert
}y\right)  $, where $g_{\mu\nu}\left(  x^{\kappa}\right)  $ is that of Bianchi
III spacetime and $f_{C}\left(  y\right)  =\sin\left(  \sqrt{\left\vert
K\right\vert }y\right)  $, where $g_{\mu\nu}$ takes the form of
Kantowski-Sachs space. Variable $\beta\left(  t\right)  $ indicates the
existence of anisotropy. When $\dot{\beta}\left(  t\right)  =0$, the
background space is that of FLRW universe.

For the line element (\ref{ch.03}) and the Action Integral (\ref{ch.01}) it
follows that the equations of motions which drive the dynamics for the
variables $\alpha\left(  t\right)  ,~\beta\left(  t\right)  $, $\phi\left(
t\right)  $ and $\psi\left(  t\right)  $ are%
\begin{equation}
e^{3\alpha}\left(  3\dot{\alpha}^{2}-\frac{3}{4}\dot{\beta}^{2}-\frac{1}%
{2}\left(  \dot{\phi}^{2}+e^{-2\kappa\phi}\dot{\psi}^{2}\right)  -V\left(
\phi\right)  \right)  -e^{\alpha-\beta}K=0, \label{ch.04}%
\end{equation}%
\begin{equation}
2\ddot{\alpha}+3\dot{\alpha}^{2}+\frac{3}{4}\dot{\beta}^{2}+\frac{1}{2}\left(
\dot{\phi}^{2}+e^{-2\kappa\phi}\dot{\psi}\right)  -V\left(  \phi\right)
-\frac{1}{3}e^{-2\alpha-\beta}K=0, \label{ch.05}%
\end{equation}%
\begin{equation}
\ddot{\beta}+3\dot{\alpha}\dot{\beta}+\frac{2}{3}e^{-2\alpha-\beta}K=0,
\label{ch.06}%
\end{equation}%
\begin{equation}
\ddot{\phi}+\kappa e^{-2\kappa\phi}\dot{\psi}^{2}+3\dot{\alpha}\dot{\phi
}+V_{,\phi}=0, \label{ch.07}%
\end{equation}%
\begin{equation}
\ddot{\psi}-2\kappa\dot{\phi}\dot{\psi}+3\dot{\alpha}\dot{\psi}=0,
\label{ch.08}%
\end{equation}
where $K=\frac{f\left(  y\right)  _{,yy}}{f\left(  y\right)  },$ is the
spatial curvature of the three-dimensional hypersurface of (\ref{ch.03}). For
Bianchi I spacetime $K=0$, for Bianchi III space is positive $K>0$ while for
the Kantowski-Sachs spacetime, $K<0$.

\section{Exact solution}

\label{sec3}

We assume the exponential potential $V\left(  \phi\right)  =V_{0}\exp\left(
-\lambda\phi\right)  $. Moreover, we observe that equation (\ref{ch.08}) is
total derivative, i.e. $\frac{d}{dt}\left(  \dot{\psi}e^{3\alpha-2\kappa\phi
}\right)  =0$. Hence the conservation law for the field equations is%
\begin{equation}
I_{0}=\dot{\psi}e^{3\alpha-2\kappa\phi}.
\end{equation}
Equation (\ref{ch.05}) can be seen as a second conservation law for the
dynamical system. In the case of a spatially flat FLRW universe, i.e.
$\dot{\beta}=0$ and $K=0$, the analytic solution of the field equation was
presented recently in \ \cite{aa5} using the Lie symmetry approach.

Hence, in order to investigate the existence of additional conservation laws,
we apply the theory of Lie symmetries. For a review on applications of the Lie
symmetry analysis in cosmology we refer the reader to \cite{sym1}. We omit the
presentation of the calculations and we give directly the results.

The dynamical system consisting of the second-order differential equations
(\ref{ch.05})-(\ref{ch.08}) for the exponential potential admits the symmetry
vectors%
\begin{equation}
X_{1}=\partial_{\psi}~~,~X_{2}=2t\partial_{t}+\frac{2}{3}\left(
\partial_{\alpha}+\partial_{\beta}\right)  +\frac{4}{\lambda}\left(
\partial_{\phi}+\kappa\psi\partial_{\psi}\right)  ~,~\text{for }\lambda\neq0
\end{equation}
with corresponding conservation laws $I_{0}$ and
\begin{equation}
I_{1}=e^{3\alpha}\left(  \dot{\beta}-4\dot{\alpha}+\frac{4}{\lambda}\left(
\dot{\phi}+\kappa e^{-2\kappa\phi}\dot{\psi}\right)  \right)  .
\end{equation}

For $\lambda=0$, that is $V\left(  \phi\right)  =V_{0}$, the admitted symmetry
vectors are the elements of the $so\left(  3\right)  $ algebra for the metric
tensor $h_{AB}$. They are
\begin{equation}
Z_{1}=\partial_{\psi},~Z_{2}=\left(  \partial_{\phi}+\kappa\psi\partial_{\psi
}\right)  ~,
\end{equation}%
\begin{equation}
Z_{3}=\psi\partial_{\phi}+\kappa\left(  \frac{\psi^{2}}{2}+\psi-\frac
{1}{2\kappa}e^{2\kappa\phi}\right)  ,
\end{equation}
with conservation laws $I_{0}$ and
\begin{equation}
\bar{I}_{2}=\left(  \dot{\phi}+\kappa e^{-2\kappa\phi}\dot{\psi}\right)
\end{equation}
and
\begin{equation}
~\bar{I}_{3}=\psi\dot{\phi}+\kappa\left(  \left(  \frac{\psi^{2}}{2}%
+\psi\right)  e^{-2\kappa\phi}-\frac{1}{2\kappa}\right)  \dot{\psi}.
\end{equation}

We focus on the case for which $\lambda\neq0$. We observe that the two
conservation laws $I_{0},I_{1}$ are not in involution, that is $\left\{
I_{0},I_{1}\right\}  \neq0$, where $\left\{  ,\right\}  $ is the Poisson
Bracket. Consequently we cannot infer the Liouville integrability of the field
equations. However, the existence of the symmetry vector $X_{2}$ indicates the
existence of invariant functions. We follow \cite{sym1} and we search for
exact solution of the form%
\begin{equation}
a\left(  t\right)  =p_{1}\ln t~,~\beta\left(  t\right)  =p_{2}\ln
t~,~\phi\left(  t\right)  =p_{3}\ln t. \label{ex.01}%
\end{equation}

We substitute into the conservation law $I_{0}$ which gives~$I_{0}%
=t^{3p_{1}-2\kappa p_{3}}\dot{\psi}$, that is
\begin{equation}
\psi\left(  t\right)  =\frac{I_{0}}{1-3p_{1}+2\kappa p_{3}}t^{1-3p_{1}+2\kappa
p_{3}}~,~3p_{1}-2\kappa p_{3}\neq1 \label{ex.02}%
\end{equation}%
\begin{equation}
\psi\left(  t\right)  =I_{0}\ln t~,~3p_{1}-2\kappa p_{3}=1. \label{ex.03}%
\end{equation}

In addition we assume that $I_{0}\neq0$, otherwise we reduce to the case of
anisotropic spaces with a quintessence field \cite{sf11}.

Let us assume now $3p_{1}-2\kappa p_{3}\neq1$, then by replacing expressions
(\ref{ex.01}), (\ref{ex.02}) in the field equations (\ref{ch.04}%
)-(\ref{ch.07}) we end with the exact solution%
\begin{align}
\alpha\left(  t\right)   &  =\frac{1}{3}\left(  1+2\frac{\kappa}{\lambda
}\right)  \ln t~,~\beta\left(  t\right)  =\frac{4}{3}\left(  1-\frac{\kappa
}{\lambda}\right)  \ln t~,~\phi\left(  t\right)  =\frac{2}{\lambda}\ln
t~\text{\ },\\
~\psi\left(  t\right)   &  =\frac{\lambda I_{0}}{2\kappa}t^{2\frac{\kappa
}{\lambda}}~,~V_{0}=\frac{\kappa}{\lambda}\left(  4+I_{0}^{2}\lambda
^{2}\right)  ~,~K=4\frac{\kappa}{\lambda}\left(  1-\frac{\kappa}{\lambda
}\right)  \text{. }%
\end{align}
with constraint equation%
\begin{equation}
\left(  4\left(  1-\kappa\lambda\right)  +\left(  2+I_{0}^{2}\right)
\lambda^{2}\right)  =0.
\end{equation}
Hence, $3p_{1}-2\kappa p_{3}=\frac{2\kappa}{\lambda}$, which means that
$2\kappa-\lambda\neq0$.

This is a new anisotropic exact solution with two scalar fields.
$\ $\ For~$K=0$, it follows that $\lambda=\kappa$. Thus $\beta\left(
t\right)  =0$, which means we end with the spatially flat FLRW spacetime. The
background spacetime is that of Bianchi III spacetime when $\kappa\left(
\lambda-\kappa\right)  >0$, while the Kantowski-Sachs metric is recovered when
$\kappa\left(  \lambda-\kappa\right)  <0$.

The anisotropic scale factor can be written as $\beta\left(  t\right)
=\frac{\lambda}{\kappa}K\ln t$, which means that the existence of a spatial
curvature indicates the existence of anisotropy. Thus, the isotropic open or
closed FLRW spacetimes as also Kasner-like universes are not provided by this
exact solution.

Indeed, this solution is the analogue of the hyperbolic inflation in the
anisotropic background space. The deceleration parameter is defined as
$q=-1-\frac{\ddot{a}}{\dot{a}^{2}}$, that is~$q\left(  t\right)
=-\frac{2\left(  \kappa-\lambda\right)  }{2\kappa+\lambda}.$ Consequently,
when the acceleration parameter is negative, $q\left(  t\right)  <0$, the
exact solution describes an accelerated solution. Thus, when $-\frac{2\left(
\kappa-\lambda\right)  }{2\kappa+\lambda}<0$, we observe that $K<0$. Hence
acceleration exists only for the Kantowski-Sachs background space. This
inflationary solution is an extension of the inflationary solution found
before for the inflaton field in Kantowski-Sachs geometry \cite{ks1}.

Finally, for the case $3p_{1}-2\kappa p_{3}=1$ by replacing expression
(\ref{ex.01}), (\ref{ex.03}) in the field equations, from (\ref{ch.06}) we
find $p_{2}=2\left(  1-p_{1}\right)  $ and $K=-3\left(  1-4p_{1}+3p_{1}%
^{2}\right)  $. Thus, from (\ref{ch.07}) it follows%
\begin{equation}
2I_{0}^{2}\kappa^{2}+\left(  1-3p_{1}\right)  t^{-1+3p_{1}}-2t^{1+3p_{1}%
+\frac{\lambda-3p_{1}\lambda}{2\kappa}}V_{0}\kappa\lambda=0. \label{ex.04}%
\end{equation}
Because we are interesting on solutions with two scalar fields, we study cases
with $I_{0}\neq0$, and $p_{1}\neq\frac{1}{3}$. Thus the polynomial equation
(\ref{ex.04}) can not be solved. Which means that there is not any anisotropic
solution of the form of (\ref{ex.01}) for $2p_{1}-2\kappa p_{3}=1$.

\section{Stability analysis}

\label{sec4}

We continue our analysis with the study of the stability properties for the
new anisotropic inflationary solution. We define the new variable $H=\dot{a}$,
and we substitute into (\ref{ch.04})-(\ref{ch.07})
\begin{align}
H  &  =\frac{\left(  1+2\frac{\kappa}{\lambda}\right)  }{3t}+\delta H\left(
t\right)  ~,~~\beta\left(  t\right)  =\frac{4}{3}\left(  1-\frac{\kappa
}{\lambda}\right)  \ln t+\delta\beta\left(  t\right)  ~,\\
\phi\left(  t\right)   &  =\frac{2}{\lambda}\ln t~+\delta\phi~,~\dot{\psi
}=I_{0}e^{-3\alpha+2\kappa\phi}~,~V_{0}=\frac{\kappa}{\lambda}\left(
4+I_{0}^{2}\lambda^{2}\right)  ~,~K=4\frac{\kappa}{\lambda}\left(
1-\frac{\kappa}{\lambda}\right)
\end{align}
and we linearize. Moreover, we perform the new change of variable, this time
for the dependent variable $t=e^{s}$. Hence we obtain the system of two linear
second-order differential equations%
\begin{align}
0  &  =3\lambda\left(  \lambda(2\kappa+\lambda)\delta\beta^{\prime\prime
}+\left(  8\kappa^{2}-6\kappa\lambda+4\lambda^{2}\right)  \delta\beta^{\prime
}+4(\lambda-\kappa)\delta\phi^{\prime}\right) \label{sw.01}\\
&  -8\kappa(5\kappa-2\lambda)(\kappa-\lambda)\delta\beta+24\kappa
(\kappa-\lambda)\delta\phi\nonumber
\end{align}
and
\begin{align}
0  &  =\lambda\left(  6(\lambda-\kappa)\delta\beta^{\prime}+\lambda
(2\kappa+\lambda)\delta\phi^{\prime\prime}+2(\kappa(2\kappa+\lambda
)+3)\delta\phi^{\prime}\right) \label{sw.02}\\
&  +12\kappa(\kappa-\lambda)\delta\beta+2\kappa\left(  (2\kappa+\lambda
)\left(  4\kappa^{2}\lambda-4\kappa-\lambda^{3}\right)  -6\right)  \delta
\phi\nonumber
\end{align}
with constraint%
\begin{equation}
\delta H\left(  s\right)  =\frac{e^{-s}\left(  \lambda\left(  (\lambda
-\kappa)\delta\beta^{\prime}+\delta\phi^{\prime}\right)  +2\kappa
(\kappa-\lambda)\delta\beta-2\kappa\delta\phi\right)  }{\lambda(2\kappa
+\lambda)}%
\end{equation}
and $\delta\beta^{\prime}=\frac{d\delta\beta}{ds}$.

\qquad The solutions of the perturbations are expressed as%
\begin{equation}
\left(  \delta\beta,\delta\phi\right)  ^{T}=%
\begin{pmatrix}
\zeta_{1} & \zeta_{2}\\
\zeta_{3} & \zeta_{4}%
\end{pmatrix}
\left(  \exp\left(  \mu_{1}\left(  \lambda,\kappa\right)  s\right)
,\exp\left(  \mu_{2}\left(  \lambda,\kappa\right)  s\right)  \right)  ^{T},
\end{equation}
where $\mu_{1}\left(  \lambda,\kappa\right)  ,~\mu_{2}\left(  \lambda
,\kappa\right)  $ are the eigenvalues for the linearized system. The
asymptotic solution is stable, when $\operatorname{Re}\left(  \mu
_{1}<0\right)  $ and $\operatorname{Re}\left(  \mu_{2}<0\right)  $.

\begin{figure}[ptb]
\centering\includegraphics[width=1\textwidth]{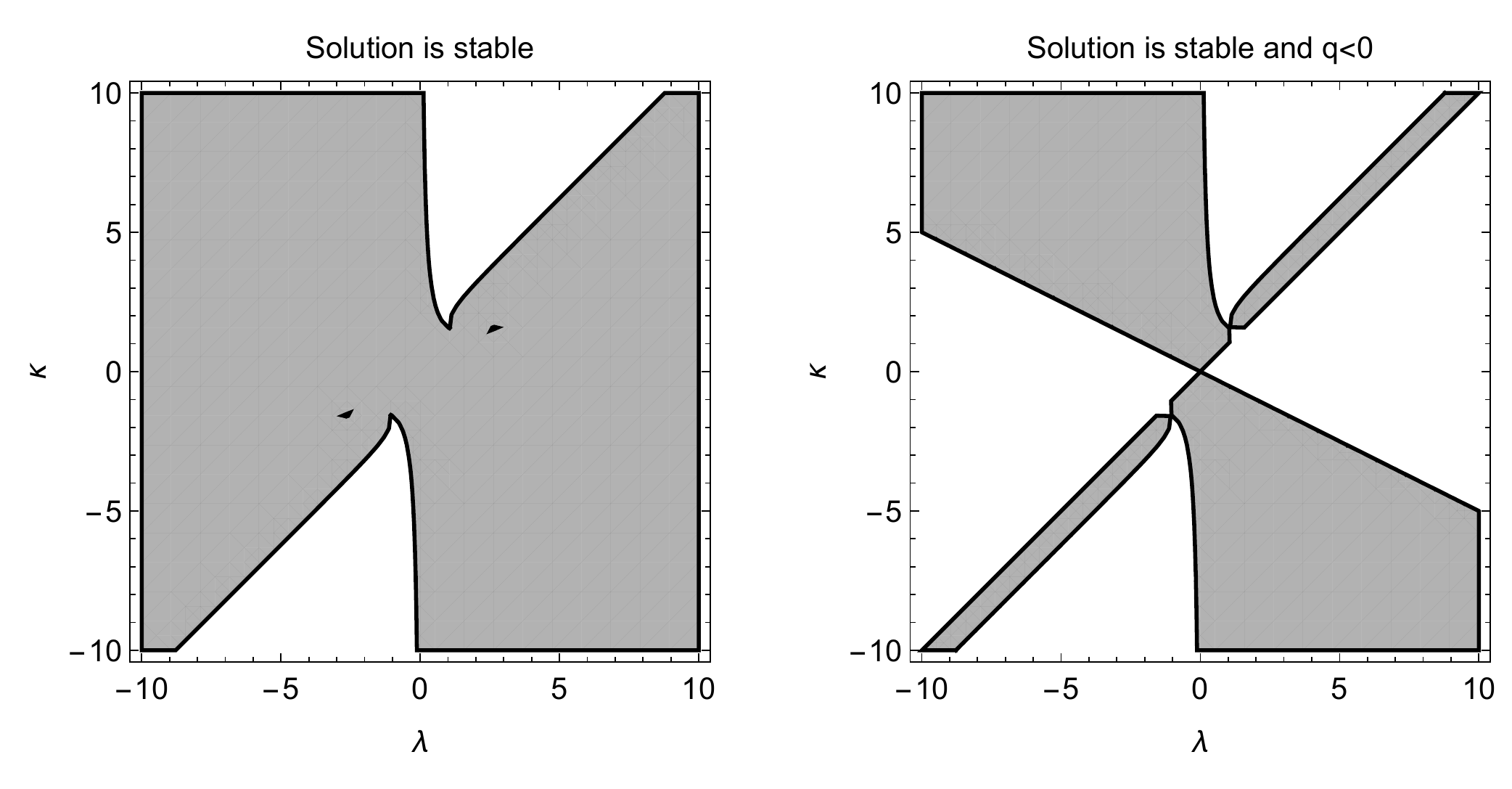}\caption{Region plot for
the eigenvalues $\mu_{1}\left(  \lambda,\kappa\right)  $, $\mu_{2}\left(
\lambda_{,}\kappa\right)  $, where the perturbations around the new anistropic
solution decays (left figure); and the perturbations around the new anistropic
solution decays and the new anisotropic solution describes an anisotopic
inflationary universe (right figure).}%
\label{fig1}%
\end{figure}

In Fig. \ref{fig1} we present the region plots for the parameters $\mu_{1}$
and $\mu_{2}$ is the space $\left(  \lambda,\kappa\right)  $, where the
perturbations decay.

\section{Conclusions}

\label{sec5}

In this work we investigated the existence of inflationary solutions on
multifield cosmology with an\ homogeneous LRS anisotropic background space. In
the context of Chiral cosmology and for the model which describes the
hyperbolic inflation in a FLRW background space, we found an anisotropic exact
solution which provides anisotropic inflation when the background spacetime
has a negative spatial curvature, that is, the physical space is described by
the Kantowski-Sachs spacetime.

For the exact solution, the anisotropic parameter and the spatial curvature
are analogues. Therefore, when the curvature term vanishes, the physical space
becomes isotropic. The method that we applied for the derivation of the exact
solution is based upon the investigation of Lie invariant functions, by
calculating the Lie symmetries for the cosmological field equations. Finally,
the stability properties for these exact solutions were studied. We found that
the inflationary anisotropic solution can be a stable solution.

In contrary to the slow-roll inflationary solution for the single scalar field
\cite{guth}, in which $\dot{\psi}=0$, and $3\dot{\alpha}\dot{\phi}%
\simeq-V_{,\phi}$, in the hyperinflation the following expressions are true
\cite{vr91}
\begin{equation}
\dot{\phi}\simeq\frac{6}{2\kappa+\lambda}\dot{\alpha}~,\label{sw.5}%
\end{equation}
which means that the evolution of the scalar field is independent on the
derivatibe of the potential. Hence, by replacing the new anisotropic solution
in (\ref{sw.5}) we find that it is true, while for the second field
$\psi\left(  t\right)  $ it holds%
\begin{equation}
e^{-2\kappa\phi}\dot{\psi}^{2}=6\left(  1-\frac{6}{2\kappa+\lambda}\right)
\dot{\alpha}^{2}-\dot{\alpha}-\frac{3}{2}\dot{\beta}^{2}--2V\left(
\phi\right)  -2e^{-2\alpha-\beta}K\label{sw.6}%
\end{equation}
where we conlclude that we have derived the analogue for the hyperinflation in
an anisotropic background space.

At this point it is important to mention that the exact solution that we found
does not provide the limit of the cosmological constant \cite{fgr}. Indeed, we
the declaration parameter is$~q\left(  t\right)  =-\frac{2\left(
\kappa-\lambda\right)  }{2\kappa+\lambda}$ and the limit for the cosmological
constant is recovered when $q\left(  t\right)  =1$, that is $\lambda=0$.
However, in our analysis we considered $\lambda\neq0$. For other forms of the
scalar field potential, it is possible to exist exact solutions which provide
the limit of the cosmological constant. Such an analysis is outside the scopus
of this work, since we focused on the exponential potential. From this results
we can infer that the Chiral model provides inflationary anisotropic solutions
which can be used as a toy model for the study of the very early universe.

Let us assume now the new anisotropic exact solution in the limit where
$\frac{\kappa}{\lambda}\simeq1+\varepsilon$, then~$\alpha\left(  t\right)
=\left(  1+\frac{2}{3}\varepsilon\right)  \ln t$ and~$\beta\left(  t\right)
=-\frac{4}{3}\varepsilon\ln t$. Hence for $\varepsilon^{2}=0$ the anisotropies
are small, and inflation can be described by the Hubble slow roll parameters
\cite{l01} $\varepsilon_{H}=-\frac{\dot{H}}{H^{2}}$,~$\eta_{H}=\frac
{\dot{\varepsilon}_{H}}{H\dot{\varepsilon}_{H}}$, from where we calculate
$\varepsilon_{H}\simeq1-\frac{2}{3}\varepsilon$ and $\eta_{H}=0$. However,
these slow roll parameters are similar with that of the exponential potential
for the inflaton field. However, because of the additional degrees of freedom,
the solution may not be always stable, thus the actual solution will be
different from the exact solution.

In a future study we plan to investigate the stability properties for the
general model as also to investigate the behaviour of the inflationary
parameters with initial conditions near to the region of the exact solution.

\end{document}